\documentclass[12pt,aps,prb,preprint,amsmath,amssymb]{revtex4}   % style for Physical Review B and AJP are similar

\usepackage{amsmath}    % need for subequations
\usepackage{graphicx}   % for figures
\usepackage{color}

\begin{document}

\title{Ephemeral properties and the illusion of microscopic particles}
\author{Massimiliano Sassoli de Bianchi}
\affiliation{Laboratorio di Autoricerca di Base, 6914 Carona, Switzerland}\date{\today}
\email{autoricerca@gmail.com}   %optional

\begin{abstract}

Founding our analysis on the Geneva-Brussels approach to quantum mechanics, we use conventional macroscopic objects as guiding examples to clarify the content of two important results of the beginning of twentieth century: Einstein-Podolsky-Rosen's reality criterion and Heisenberg's uncertainty principle. We then use them in combination to show that our widespread belief in the existence of microscopic particles is only the result of a cognitive illusion, as microscopic particles are not particles, but are instead the ephemeral spatial and local manifestations of non-spatial and non-local entities.


\end{abstract}

\maketitle

\section{Introduction}
\label{intro}

A majority of physicists, be them theoreticians or experimentalists, researchers or teachers, still believe that the basic building blocks of our physical reality are the so-called \emph{elementary particles} (quarks, leptons and gauge bosons). This however is a false belief, based on our macroscopic prejudice about the microscopic reality, as since the early days of quantum mechanics there has been strong evidence that: \emph{microscopic particles do not exist}. 

A number of authors have pointed out, as years go, the incorrectness of the corpuscular vision and the inevitable conceptual confusion that a particle ontology can only produce. A prominent example is Nobel laureate Steven Weinberg, whose statement that quantum fields form our ``essential reality'' and that particles are reduced to a mere epiphenomenon, is often quoted~\cite{Weinberg}. But despite all these didactical efforts, we have to admit that, conceptually speaking, the particle mythology still persists. 

The main purpose of the present article is to contribute to this longstanding and ongoing effort of reeducation of physicists' (and philosophers of science) worldview, by presenting what we believe is a clear argument in favor of a non-particle ontology. 

To this end, in Sec.~\ref{sec:1}, we start by making precise what we mean exactly by the concepts of \emph{microscopic} and \emph{particle}. Then, in Sec.~\ref{sec:2}, we consider the notion of \emph{element of reality}, as firstly introduced by Einstein Podolsky and Rosen in their 1935 paper~\cite{Einstein}, and use it to present and clarify the content of a more complete \emph{existence criterion} (EC), which was introduced by Constantin Piron~\cite{Piron1, Piron2, Piron3} as a key ingredient of the so-called Geneva-Brussels (realistic) approach to quantum mechanics. In Sec.~\ref{sec:3}, we recall the conceptual content of Heisenberg uncertainty principle (HUP) and its relation with the important notion of \emph{experimental incompatibility}, exploiting for this Aerts' celebrated example of a macroscopic wooden entity~\cite{Aerts1}. Then, combining the HUP with the EC, we prove in Sec.~\ref{sec:4} the inexistence of microscopic particles, as a direct consequence of their \emph{lack of spatiality}. 

Let us emphasize from the outset that the arguments we present in this work are not new in themselves, as they were already considered by Diederik Aerts, in different ways, in a number of fundamental papers, where the important notion of non-spatiality of quantum entities has been thoroughly investigated~\cite{Aerts2, Aerts3, Aerts4}. However, our presentation will be more focused on the direct consequence of non-spatiality for what concerns the validity of the (still) widespread notion of ``microscopic particle.'' 

Also, our goal is to further illustrate (and consequently explain) the reasons for such an observed non-spatiality of quantum entities, introducing for this purpose the notion of \emph{ephemerality} of a quantum property. We do this following Aerts' tradition of inventing creative examples of conventional macroscopic objects exhibiting non-classical behavior, where all the quantum mystery is, in a sense, under our eyes. 

More precisely, in Sec.~\ref{sec:5} we present what we believe is the first model of a  macroscopic entity (a simple uncooked Italian spaghetti!) exhibiting non-classical and non-compatible properties. Thanks to this example, which complements the already rich collection of macroscopic models provided by Aerts, we can gain a better understanding of the relationship between the ephemeral character of certain quantum properties and their observed non-spatiality. 

In Sec.~\ref{sec:6}, we carry on with our didactical effort by analyzing EPR's experiment, showing that there is another fundamental attribute that the presumed microscopic particles fail to possess: individuality. 

Finally, In Sec.~\ref{sec:7}, we briefly turn our attention to the equally misleading concept of \emph{quantum field} and present a symbolic ``mea culpa'', while in  Sec.~\ref{sec:8} we offer some concluding remarks.

\section{Microscopic Particles}
\label{sec:1}

An entity will be said to be \emph{microscopic}, in the sense used in this article, if Heisenberg's uncertainty principle (HUP) is relevant for the description of its kinematic observables, like position and momentum. 

Concerning the concept of \emph{particle}, we can observe that there are essentially two meanings that are today associated with it. The original, etymological meaning, is that a particle is a \emph{small part} of a bigger composite system, i.e., a small subsystem. The second, more general and nowadays more common meaning, is that a particle is simply a \emph{small localized physical system}, a corpuscle, a small body which doesn't necessarily need to be part of a bigger composite system. And this is also the meaning we will adopt in the following. 

Undoubtedly, many different \emph{properties} can be associated to the concept of particle, or corpuscle. However, for our purpose it will be sufficient to focus on one of the most fundamental: \emph{spatiality}. If a physical entity is a particle, then, in every moment, it must possess a specific location in space, characterizable by a position (for instance of its center of mass) with reference to a given coordinate system. In other terms, it has to \emph{exist somewhere in our three-dimensional Euclidean physical space}.

\section{Existence Criterion}
\label{sec:2}

Finally, let us define the concept of \emph{reality} or, which is equivalent, of \emph{existence}, as in the common understanding of these two terms something is said to be real if and only if it exists. What we need is a criterion to be used to discriminate between what exists (and therefore is part of reality) and what doesn't exist (and therefore is only potentially, but not actually, part of reality). 

Such a criterion was provided by Einstein and his two collaborators, Podolsky and Rosen (abbreviated as EPR in the sequel), who in their famous article of 1935 enounced the following reality criterion (emphasis is our)~\cite{Einstein}: ``If, without in any way disturbing a system, we can \emph{predict with certainty} [...] the value of a physical quantity, then there \emph{exist} an \emph{element of physical reality} corresponding to this physical quantity.'' 

What EPR have clearly recognized is that our description of reality is essentially based on our dependable predictions about it. However, they remained rather cautious regarding their criterion, as they also wrote~\cite{Einstein}: ``It seems to us that this criterion, while far from exhausting all possible ways of recognizing a physical reality, at least provides us with one such way, whenever the conditions set down in it occur. Regarded not as a necessary, but merely as a sufficient, condition for reality, this criterion is in agreement with classical as well as quantum-mechanical ideas of reality.''

Despite their warning, EPR didn't offer a single counter example of what would be the nature of an element of physical reality not subject to their criterion. In other terms, although they assume, very prudently, that their criterion is only sufficient, they present no reasons as to why it shouldn't be considered also necessary. 

An important point to emphasize is that when EPR write ``if [...] we can predict with certainty'', what one  should understand is: ``if we can \emph{in principle} predict with certainty''. In fact, the crucial point is not if we do possess in practice all the relevant information allowing us to make a reliable prediction, but if this information is available somewhere in the universe (although maybe dispersed who know where), so that a being of sufficient power and intelligence could \emph{in principle} access it. 

In their article, EPR didn't mention explicitly this subtle point, and this author is not aware if there have been discussions of Einstein and collaborators, after their 1935 paper, where the ``in principle'' issue (which is at the core of the relationship between the \emph{ontic} and \emph{epistemic} perspective in our description of reality) was mentioned and clarified. However, what is certain is that the subject matter was subsequently re-evaluated by Piron, who cleverly incorporated these ideas in his elaboration of the key notions of \emph{actual property} and \emph{true test}~\cite{Piron1, Piron2, Piron3}. 

There is indeed a very well developed approach to the foundations of quantum mechanics, generally referred to as the Geneva-Brussels approach (which counts within its founders J. M. Jauch, C. Piron, D. Aerts and others; see for instance Refs.~\onlinecite{Piron1, Piron2, Piron3, Aerts1, Aerts2, Aerts3, Aerts4, Aerts5, Aerts6, Christiaens} and the references cited therein), where the notion of \emph{element of reality} introduced by Einstein Podolsky and Rosen has been further developed in a very detailed and nuanced way. This has been done also with the aim of putting forward a very specific and very complete \emph{Existence Criterion}, which is the following:\\

\noindent\textbf{Definition} (Existence Criterion): \emph{If, without in any way disturbing the physical entity under consideration, it is in principle possible to predict with certainty the outcome of a given experimental test, then the property associated to that test is an actual (existing) property of the entity. Conversely, if the property of a physical entity is actual, then without in any way disturbing it, it is in principle possible to predict with certainty the outcome of one of the associated experimental tests.}\\

According to the above Existence Criterion (EC), a property is actual \emph{if and only if}, should one decide to perform one of the (equivalent) tests that define it, the expected result would be \emph{certain in advance}. This means that the entity \emph{has} the property in question before the test is done, and in fact even before one has chosen to do it. And this is the reason why one is allowed to say that the property is an element of reality, existing independently from our observation. 

On the other hand, if we cannot apply the above EC, that is, if we cannot, \emph{not even in principle}, predict the outcome of a test defining the property in question, then we must conclude that the entity under consideration doesn't possess that property, i.e., that the property is not an actual (existing) one. 

This conclusion is correct provided the prediction cannot be made \emph{even in principle}. In fact, in most experimental situations we simply do not possess a complete knowledge of the entity, and therefore haven't access to all its actual properties. But when we do possess a \emph{complete knowledge} of the entity, then, by definition, we are able to predict with certainty all that is predictable about it, and therefore what cannot be predicted is by definition a non-existing (potential, uncreated) aspect of reality. To put it in different terms, to have a complete knowledge about an entity is to have full knowledge about its \emph{state}, which represents what the entity \emph{is}, its \emph{reality} at a given time, given by the set of all its actual properties (see the above mentioned references of Piron and Aerts, as well as the interesting conceptual analysis in Ref.~\onlinecite{Smets}).

\section{Incompatibility}
\label{sec:3}

Having clarified the three concepts mentioned in the introduction, to present our argument we need to turn to our last ingredient: HUP. This principle (which strictly speaking is not one) expresses the fact that on a quantum entity one cannot simultaneously extract information about two non-commuting observables (for instance position and momentum); the non-commutability of the operators representing these observables being the mathematical counterpart of the incompatibility of the corresponding experimental procedures that are used to define and measure them.

HUP poses a fundamental interpretational problem: if two observables associated to a quantum entity are mutually incompatible, and therefore entertain an uncertainty relation, does this mean that the entity in question cannot \emph{jointly possess} specific values relevant to these two observables, or just that the experimenter cannot jointly access these values, because of the incompatibility of the corresponding experimental procedures? In other terms, is the HUP a statement about the \emph{non-existence} of certain non-commuting observables, and corresponding properties, or is it just a statement about our limitations in jointly measuring them, independently from their existence? 

To clarify the content of fundamental interrogatives of this sort, the notion of compatibility has been analyzed in depth by Aerts, within the Geneva-Brussels approach. In particular, Aerts introduced the analysis of a simple macroscopic system - a small piece of wood - for which all the mystery of experimental incompatibility is under our eyes~\cite{Aerts1}. Let us consider it once more.

A small piece of wood is a physical entity possessing a number of properties, like for instance the one of ``burning well'', or of ``floating on water''. To observe the ``burning well'' property, we need to put the piece of wood on fire, whereas to observe the ``floating on water'' property, we need to immerge it in the water. However, these two experimental tests are clearly incompatible, as a wet piece of wood will not any more burn well, and a burned piece of wood will not in general float. In other terms, we cannot jointly observe the ``burning well'' and ``floating on water'' properties of the piece of wood entity. 

But, independently of their experimental incompatibility, the piece of wood does jointly actually possess these two properties, as is clear from the fact that we can predict in advance, with certainty, that whatever test we may chose to perform, the outcome will be a ``yes'' (this is the essence of what is technically called a \emph{product test}~\cite{Piron3, Aerts1}).

If we use the piece of wood entity as a paradigmatic example of experimental incompatibility, we may be tempted to conclude that nothing prevent a quantum entity to jointly possess well-defined values associated to incompatible observables, as one thing is to simultaneously possess two properties (like a specific position and momentum), and another one is to be able to jointly access their value experimentally. As we shall see, this conclusion is however wrong for quantum observables.

\section{Non-spatiality}
\label{sec:4}

A \emph{microscopic particle}, as we said, is a physical entity possessing at least the attribute of spatiality and whose kinematic observables are subject to HUP. What we will now show is that such a definition is an empty one, as no entity of this kind can exist (i.e., be real). 

As mentioned in the introduction, the notion of non-spatiality, in relation to quantum entities, has already been investigated by Aerts, in a number of publications~\cite{Aerts2, Aerts3, Aerts4} and we don't pretend here being particularly innovative in our discussion, which is just another way to present a well-known argument in what we hope is a conceptually clear and precise way.

Spatiality is the attribute of ``being always present somewhere in the three-dimensional Euclidean physical space, in the course of one's existence.'' What's important is not if we know in practical terms the locations where a particle-entity actually is and will be, but if we can know in principle its positions, assuming we would have a \emph{full knowledge of its state}. 

Let us start considering the case of a \emph{macroscopic particle}, that no doubts possesses the attribute of spatiality. In fact, seeing that it is not subject to HUP, we can measure its position at a given instant, and simultaneously its momentum; and once we know their values at a given moment, we can solve Newton's equations of motion (whose solutions can be proven to exist) and use them to predict with certainty, without disturbing the particle, all its future spatial locations. 

Based on the certainty of these predictions, we can then conclude, thanks to the EC, that in every moment the macroscopic particle is somewhere in space, i.e., that it exists in space. 

When instead of a macroscopic body we consider a (hypothetical) microscopic corpuscle, what changes is that HUP now applies, so that when we measure its spatial localization, we destroy the possibility of knowing its momentum value, and vice versa. Based on our macroscopic prejudice on reality, we may nevertheless cling to the idea that the microscopic particle does \emph{actually} possess a position and a momentum, but because of the inevitable disturbance introduced by our measurements, we simply cannot discover, in practical terms, their simultaneous values (in the same way as we cannot simultaneously observe the ``burning well'' and ``floating on water'' properties of the macroscopic wood entity, despite they clearly jointly exist). This belief is however inconsistent, as revealed by the following Proposition.\\

\noindent\textbf{Proposition} (Non-Spatiality): \emph{Let $S$ be a physical entity, whose kinematic observables obey HUP. If the EC is valid, then $S$ is non-spatial.}\\ 

\noindent\textsl{Proof}: Even with a full knowledge of the state of $S$, HUP prevents us to simultaneously determine its position and momentum. Therefore, we cannot determine (by solving the Lagrangian's or Hamiltonian's equations), even in principle, how the position of $S$ will vary in time and predict with certainty its future locations. Using the EC, we can thus conclude that $S$ doesn't actually possess the property of being somewhere in physical space, so that whatever $S$ is, it is a \emph{non-spatial} entity.\\

It immediately follows from the above result that \emph{microscopic particles do not exist}, as to be particles microscopic entities must be present in our three-dimensional physical space. But since they don't, their corpuscular nature is just the result of a cognitive illusion.

Let us stress, for sake of clarity, that this conclusion is mandatory only if we agree that the EC correctly draws the boundary between existence and non-existence. Nonetheless, adopting an alternative criterion that would for instance assert the existence of microscopic corpuscular entities in themselves, independently of our possibility of predicting, neither in principle, their positions in space, would clearly be a choice of metaphysical nature, falling beyond the strict scope of scientific investigation.

This is the stance taken by the de Broglie-Bohm theory, where microscopic particles having well defined locations in the three-dimensional Euclidean space are assumed \emph{ad hoc} to exist, independently of our possibility of predicting, even in principle, their actual positions; this because of the presence of an hypothetical causal field, whose existence also needs to be postulated \emph{ad hoc}, which would manifest at a subquantum level of reality and whose random fluctuations, once averaged over time, would yield the ordinary quantum wave function. 

We will not comment here any further the metaphysical particle ontology of the de Broglie-Bohm theory, which is known to present serious interpretational problems when one attempts to describe more than a single quantum entity (see for instance the discussion in Ref.~\onlinecite{Aerts3}).

\section{Ephemeral Properties}
\label{sec:5}

Aerts and his collaborators in Brussels have introduced an approach which they refer to as the \emph{hidden measurement approach}~\cite{Aerts7, Aerts3, Aerts4}, where the situation of \emph{non-existence} of position and other observables of a quantum entity, despite our complete knowledge of its state, is explained by introducing the hypothesis that these quantities are literally \emph{created} by the very experiments we use to operationally define and measure them. The naming ``hidden measurement'' refers to the fact that the measurements that provoke this creation in the case of quantum observables are hidden, in the sense that macroscopically indistinguishable measurements can in fact microscopically be different (due to the effect of these hidden measurements), which would then also explain the presence of an irreducible probability.

An important role in Aerts' analysis of the creation aspect of a quantum measurement has been played by his remarkable \emph{quantum machine entity}, a macroscopic mechanical model the description of which is isomorphic to the description of the spin of a spin $1/2$ ``particle''. Indeed, thanks to Aerts' macroscopic model, it is possible to explicitly observe how a measurement process can be responsible for the creation of a new unpredictable state for the quantum machine entity, that was inexistent prior to the measurement, and this despite our complete knowledge of its state prior to the measurement. 

Without going into the details, let us briefly recall what are the basic elements constituting Aert's quantum machine entity (for a complete description and analysis of the model, we refer the interested reader to Refs.~\onlinecite{Aerts3, Aerts4} and references cited therein). Roughly speaking, a quantum machine entity is a point particle localized inside a three-dimensional Euclidean sphere, the different possible states of which are the different places the particle can occupy inside of it. The particularity and cleverness of the model resides in the way experiments are designed. Indeed, to observe the state of the entity the experimental protocol is to use a sticky elastic band that is stripped between two opposite points of the sphere (each couple of points defining a different experiment); then, one simply let the point particle fall from its original location orthogonally onto the elastic and stick to it. At this point the elastic breaks, at some unpredictable point and the point particle, which is attached to one of the two pieces of it, is pulled to one of the two opposite end points, which is the outcome of the experiment, i.e., the state that is acquired by the quantum machine entity as a result of the measurement.

We can observe that in Aerts' model the entity in itself is quite conventional (a classical point particle) whereas what is quite unusual is the measurement process, which exploits the breakability of an elastic band. What is extremely interesting in the model is that it fully illustrates a situation where the outcomes of the experiments can only be predicted in probabilistic terms, and this not because of a lack of knowledge of the state of the system, but because of a lack of knowledge about the measurement process, namely where exactly the elastic band breaks during its execution. 

Now, although the quantum machine entity allows for a full modelization of a spin $1/2$, it doesn't illustrate an important general feature of quantum properties: their \emph{ephemerality}. Indeed, the position for instance of a microscopic entity, like an electron, is ephemeral in the sense that, in general, not only one cannot predict its value prior to the measurement, but neither can one do it a finite time following it, however small such a finite time is. The reason for this, as we emphasized with the proposition of the previous section, is that not only such a position is in fact  inexistent prior to the measurement, but  it also ceases to exist immediately following it. 

The above conclusion is mandatory if one accepts the general validity of the EC and HUP. However, regardless of one's personal predisposition in believing in the ephemerality of quantum properties as a consequence of the non-spatiality of the corresponding quantum entities, it would certainly be useful to dispose of a macroscopic model ``\`a la Aerts'', that would help us in guiding our intuition. 

This is what we will do now, by presenting a specific example of a macroscopic entity displaying not only the typical ephemerality of quantum properties, but their incompatibility as well. The prospect of such a modelization was already contained, as a possibility, in the general framework developed by Aerts in his \emph{hidden measurement approach}~\cite{Aerts7, Aerts4} and, more generally, in his \emph{creation discovery view}~\cite{Aerts3, Aerts4}. However, as far as we know such an explicitly model hasn't been worked out so far.

The macroscopic entity (or system) we consider is a conventional \emph{uncooked Italian spaghetti}, which can either be whole or broken into fragments. We want to measure (i.e., observe) the ``left handedness'' property of the spaghetti, which we define by the following test: bend it until it breaks, if the longest fragment remains in the left hand the answer is ``yes,'' otherwise ``no;'' and if the spaghetti is already broken, simply do the experiment using the longest fragment. 

Obviously, we cannot determine in advance if a spaghetti is left-handed (or alternatively right-handed), and this is not imputable to our lack of knowledge about its state. Even with a full knowledge of all the spaghetti's actual properties, down to the molecular level, we could not predict the outcome of the experiment, as the ``left handedness'' property is \emph{created during the test itself} (i.e., during its observation), according to the specific points where the spaghetti-entity breaks, that depend on a number of fluctuating factors that are completely out of our control (similarly to what happens with the breaking of the elastic band in the quantum machine model).

In other terms, although we may possess a complete knowledge of the state of the spaghetti-entity, we don't have any control of the interaction between the entity and the apparatus (our hands). And because of that, the better we can do is to predict the outcome in probabilistic terms. 

To repeat what has been put forward many times by Aerts in his \emph{hidden measurement approach}~\cite{Aerts7, Aerts4}, what distinguish quantum from classical probabilities could be nothing but the fact that the latter describe our lack of knowledge about what already exists, whereas the former describe our lack of knowledge about what is contextually brought into existence by our experiments.

Clearly, the left-handedness (or right-handedness) of a spaghetti string is an \emph{ephemeral property} that comes into existence only in the very moment the test is executed, but also ceases to exist immediately after it, as if we want to repeat the test (using the longest fragment), the outcome is again unpredictable. 

An important feature of our spaghetti model is that it highlights the fact that properties of the left-handedness kind, despite their ephemeral character, can nevertheless entertain incompatibility relations with other ephemeral properties. To see this, consider another property of the uncooked spaghetti, that we call ``solidity'' and define as follows: let the spaghetti fall from your hand to the floor, if it doesn't break the answer is ``yes,'' otherwise ``no;'' and if the spaghetti is already broken, simply do the experiment using the longest fragment. 

Similarly to the ``left-handedness'' property, the ``solidity'' property is also created during the very execution of the test used to define it, and is also of an ephemeral nature. Nevertheless, it clearly entertains an incompatibility relation with the ``left-handedness'' property, seeing that testing the latter considerably increases the probability for the test of the former to give a positive answer (as is clear from the fact that the shortest the spaghetti fragment the less easily it will break when falling to the floor). 

Our spaghetti model is, in a sense, complementary to Aerts' piece of wood model. Indeed, both systems exhibit incompatible properties, but in the case of the piece of wood these properties (``burning well'' and ``floating on water'') are classical, in the sense that they are stably existing in our three-dimensional space (which is the space in which these properties are defined and made manifest by means of the corresponding experimental tests), whereas in the case of the spaghetti (``left handedness'' and ``solidity'') they are non-classical, in the sense that they are only ephemerically and unpredictably brought into existence into it by the very experimental test that define them. 

Thanks to these two paradigmatic examples, we can clearly observe that incompatibility and ephemerality are independent concepts. Indeed, as the piece of wood model shows, two properties can stably exist although being incompatible, but also, as the spaghetti model shows, they can as well only ephemerically exist and nevertheless still being incompatible. This indicates that the ephemeral character of a non-classical property has more to do with the way the property itself is defined (i.e., tested) as to the fact that it may or not entertain incompatibility relations with other properties. 

This remark is relevant in view of the fact that the non-spatiality argument presented in the proposition of the previous section uses HUP as its main ingredient, i.e., the existence of a relation of incompatibility between position and momentum. Therefore, one may be tempted to conclude that it is the very existence of such an experimental incompatibility which is at the origin of the observed non-spatiality (and consequent ephemerality of the spatial position) of microscopic entities. However, considering the piece of wood example, we see that incompatibility is not a sufficient condition for non-spatiality (or, more generally, ephemerality), and considering the spaghetti example we also see that incompatibility is neither a necessary condition for it, as is clear from the fact that the ephemeral character of the ``left handedness'' or ``solidity'' properties is built-in in the very definition of them, independently of the compatible or incompatible nature of their relationships with other properties. 

In other terms, although HUP can be conveniently used to establish the non-spatiality of microscopic quantum entities, this doesn't mean it is a fundamental ingredient of it. In fact, the importance of HUP in this context is to be linked to the specific structure of the Lagrangian (or Hamiltonian) formulation, that requires position and momentum to be \emph{independent} variables for the unique determination of the solution of the equation of motion. But, as the piece of wood and spaghetti models show, there is no logical necessity to resort to HUP (i.e., to incompatibility), to establish the non-spatiality of quantum entities. 

In fact, as it has been many times pointed out by Aerts~\cite{Aerts3, Aerts4}, the widespread quantum effect called \emph{non-locality}, that manifests for instance in Rauch's neutron interferometry experiments, Wheeler's delayed choice experiments, or Aspect's experiments with entangled pairs, when closely analyzed, can in fact be considered as a clear and direct signature of non-spatiality.

\section{Individuality}
\label{sec:6}

In the previous sections we have considered the piece of wood and spaghetti models as paradigmatic guiding examples to observe that entities can have classical properties, which exist independently of our observations, and quantum (or quantum-like) properties that we can ephemerally bring into existence by means of our experiments. As we have seen, these quantum properties being extremely short-lived, they cannot be used to characterize the identity of an entity, as for this one needs to find invariant properties (i.e., attributes) whose actuality must be predictable with certainty and remain such for a sufficiently long time. And, as we emphasized, spatiality is a typical property that an entity needs to invariantly possess to be identified as a particle. 

If the entity is microscopic, the property of ``being spatial'' can be actualized just for a moment. However, \emph{manifesting a single position just for one moment doesn't mean creating a particle with an entire trajectory!}

In this section we want to further strengthen the argumentation regarding the inadequacy of the corpuscular vision, by explaining that, in addition to spatiality, the hypothetical microscopic particles lack another fundamental attribute: \emph{individuality}.

It is important to recall that EPR enounced their reality criterion with the intent of disproving the limitations imposed by HUP and show that it was possible to ascribe a joint reality to the position and momentum of a microscopic particle. To explain the essence of their reasoning, let us come back for a moment to Aerts' piece of wood entity. As we observed, the ``burning well'' and ``floating on water'' properties are experimentally incompatible, and this means we cannot jointly test them (i.e., observe them) without affecting their respective outcomes. The situation however would be different if we could dispose of two identical pieces of wood, i.e., two identical \emph{separated} entities, on which we could jointly perform one of the two experiments. In this way, they would clearly not mutually affect one another anymore, and we could collect the expected ``yes'' outcomes from both of them. 

This is exactly the idea behind EPR's celebrated \emph{gedankenexperimente}: instead of trying to jointly measure the position and momentum on a same microscopic particle, one can  instead consider two (hypothetical) identical and separated microscopic particles, and perform on each of them one of the two tests, whose outcomes can then be jointly collected, thus demonstrating the possibility of by-passing the position-momentum Heisenberg's uncertainty relation. 

More specifically, what EPR showed in their 1935 paper is that if two identical particles are prepared in a specific (product) state, and then are left to interact, they form a so-called (non-product) entangled state, in which the (hypothetical) properties possessed by the two individual particles  are strongly correlated. But since with time the two particles become more and more separated in spatial terms, according to EPR they also become more and more separated in experimental terms, as no influence can travel instantaneously the spatial distance between them and produce a so-called ``spooky action-at-distance.'' 

According to this line of reasoning, it becomes possible, at least in principle, to measure the momentum of, say, the first particle, then use the outcome of the measurement to predict with certainty the momentum of the second one (thanks to the correlation induced by the entanglement), without in any way disturbing the latter. Then, one is still free to perform on the second particle a measurement of its position, thus showing the possibility of simultaneously determining both observables. And since there would be a contradiction between the conclusion reached by EPR's subtle reasoning, and the predictions of HUP (and more generally of QM), that's why historically this situation has been referred to as the EPR's \emph{paradox}, and that's also why EPR concluded that quantum mechanics is an \emph{incomplete} theory, as it cannot account for all the actual properties, i.e., for all the possible elements of reality. 

However, as we today know thanks to the analysis of Bell~\cite{Bell1, Bell2} and the historical experiments with entangled pairs of Aspect et al.~\cite{Aspect1, Aspect2} (and the many others that since then followed) the false prejudice of EPR was to believe that putting a sufficient spatial distance between two (hypothetical) microscopic particle-entities would automatically guarantee their experimental separation. In other terms, they mixed up the concept of \emph{experimental separation} with the one of \emph{spatial separation} (or absence of interaction). 

In fact, what the experimental violation of Bell's inequalities has clearly shown is that QM correctly describes the state of the system formed by the two spatially separated microscopic entities emerging from the interaction by means of an entangled (non-product) state, so that the system cannot be understood as the sum of two (experimentally) separated parts (each one described by its own specific state), but as a single genuine whole entity. 

This shows that if microscopic particles cannot exist, it is not only because they aren't present in space, as we have shown in the previous section, but also because they generally don't possess the essential attribute of \emph{individuality}, i.e., the property of existing as separate entities.  And such an attribute, like the one of spatiality, is certainly one we cannot relax if we want to preserve the meaningfulness of the very concept of particle. 

It is worth mentioning that the experimental violation of Bell's inequalities didn't provide a solution of the EPR paradox per se. Such a solution only became possible in the eighties, following Aerts' deep analysis of the EPR situation, which he presented in his Doctoral Thesis and then further developed in a number of publications. In these works, he pointed out a very subtle and crucial point: that standard QM is in fact unable, structurally speaking, to describe systems made of separated entities~\cite{Aerts1, Aerts5} (the concept of \emph{separation} is here to be understood in the experimental sense, i.e., in the sense that performing experiments on one of the entities doesn't affect the state of the others, and vice versa). 

As stressed by Aerts, this is a fundamental (and usually misunderstood) point, as EPR, in their (ex-absurdum) reasoning, precisely assumed, on the contrary, that QM was perfectly able to describe the situation of two entities that become separated (in the experimental sense) as they fly apart in space. But this premise was proven by Aerts' analysis to be false. So, although EPR conclusion about the incompleteness of QM was correct, it wasn't for the reasons advocated in their 1935 paper. If QM is incomplete, it is because it fails to correctly describe separated physical entities~\cite{Aerts1, Aerts5}  (see also the discussion in Ref.~\onlinecite{Christiaens}).

\section{Entities}
\label{sec:7}

Following the Geneva-Brussel approach to QM, we have considered in the previous sections two important arguments in favor of a (microscopic) non-particle ontology: the prominent non-spatiality and non-individuality of quantum entities. This implies that QM, in its usual formulation and interpretation, becomes a wrong theory when it pretends giving a proper meaning to the illusory concept of ``individual microscopic particle.'' And, each time it does so, it automatically gives rise to a number of unnecessary paradoxes. 

It should be mentioned that, in addition to spatiality and individuality, many others (interrelated) attributes can also be associated to a physical entity, to properly characterize its corpuscular nature, like for instance: distinguishability, materiality, impenetrability, movement, shape, and many others as well.

The definition of many of these attributes can certainly be relaxed, without necessarily renouncing to the very concept of particle. For example, we may still agree to call an entity a particle even though it is not rigorously impenetrable, and could for instance be passed through by other entities, though not by all entities. In the same way, without giving up the concept of particle, we may replace the requirement of possessing a certain amount of matter with the more general one of possessing a certain amount of energy, as mass, in ultimate analysis, is internal energy, and one can imagine particles having transformed all their internal energy in kinetic energy, thus moving at the maximum allowed speed (as it is the case for the photon-pseudoparticle, in case its mass would be strictly zero).

On the other hand, there is to our opinion little room to relax the attributes of spatiality and individuality, without renouncing to the very concept of particle. If an entity is not in every moment present somewhere in space, in the course of its existence, and cannot be separated from other entities, then it cannot be considered a particle, at least not in the sense we usually understand this concept. 

Said this, we have nevertheless to acknowledge that the hypothesis of microscopic particles, although incorrect, has proved to be an incredibly fertile idea. Quoting for instance Feynman, from the first pages of his famous lectures in physics~\cite{ Feynman1}: ``If, in some cataclysm, all of scientific knowledge were to be destroyed, and only one sentence passed on to the next generations of creatures, what statement would contain the most information in the fewest words? I believe it is the atomic hypothesis (or the atomic fact, or whatever you wish to call it) that all things are made of atoms-little particles that move around in perpetual motion, attracting each other when they are a little distance apart, but repelling upon being squeezed into one another. In that one sentence, you will see, there is an enormous amount of information about the world, if just a little imagination and thinking are applied.''

Although Feynman was certainly wrong in pretending that the atomic hypothesis is a fact, he was undoubtedly correct in considering that atomism, and more generally reductionism, has been and still is a source of great explicative power. Physical systems are not made of atoms, or of elementary particles, but doing \emph{as if} they are can certainly be useful in many circumstances. Yet, we must not forget the ``as if'', because then the risk is to miss all the more advanced explanations and predictions that lie beyond our incomplete corpuscular vision, which is just a lucky conceptual analogy extrapolated from our experience of the macro world.

There is however another important point we don't have to forget, when discussing QM. Quoting G. Preparata (Ref.~\onlinecite{Preparata}, page 63): ``[...] quantum mechanics is a rigorous consequence of the corresponding quantum field theory in the limit of `infinite dilution', where the world is only populated by a small, finite number of quanta''. 

In other terms, QM is just an approximation of a more general and advanced theory, called \emph{quantum field theory} (QFT), and, as its name indicates, the basic ingredients of QFT are not particles, but quantum fields! 

When quantum fields interact together, they do it locally, exchanging quanta of energy, and when they do so they leave visible traces in our three-dimensional physical space, for instance in the form of little spots on a screen. These traces are easy to be mistaken for the traces that would be left by some hypothetical microscopic corpuscles with a proper kinematics and dynamics. This however is only a cognitive illusion, as these traces are just the consequence of the fact that quantum fields, contrary to classical ones, can only exchange their energy in a discontinuous way, that is, in small packets, or bundles, or quanta.   

Coming back to Weinberg's quote mentioned in the Introduction, that quantum fields form our ``essential reality'' and that particles are reduced to a mere epiphenomenon, we may then ask if, conceptually speaking, it would be sufficient to substitute the notion of ``particle'' by the notion of ``field'' to solve all interpretational problems. 

Some physicists do indeed strongly defend the point of view that most of our difficulties in understanding QM are the result of our bad habits of still thinking to quantum phenomena with classical corpuscular ideas, and that the way out of this impasse is just to replace the outdated concept of ``particle'' by the one of ``field,'' something that is believed could be easily done as from the level of our introductory courses of quantum physics, thus offering to the students a more unified, realistic and less paradoxical view of the physical reality (see for instance Refs.~\onlinecite{Hobson, Scandal1, Scandal2, Scandal3, Scandal4} for some recent interesting discussions on this topic).

Surely, the notion of ``field'' is, in a sense, a more suitable notion than the one of ``particle,'' to coherently describe the reality of our microscopic world. Indeed, it overcomes the problem of having to think about quanta as individual tiny particles moving in empty space, which instead can be understood as the traces left by quantum fields as they interact with our instruments; traces that, necessarily, must be present in our ordinary physical space.

However, at a closer look, we must admit that the substitution of the notion of (microscopic) ``quantum particle'' by the notion of ``quantum field'' has the sole effect of shifting the problem. Indeed, a field is also an entity defined in space, with specific properties in every point of it (like for instance force vectors), and therefore even such a notion is unable to conveniently describe the typical non-spatial nature of a quantum entity and so avoid all paradoxes. 

Contrary to particles, fields are non-local, ``spread out'' entities, but they are still spatial entities. The only difference is that a particle is imagined to have one specific, almost point like, location at any moment of time, whereas a field is imagined to have a spread out location in space, at any moment of time (similarly to a wave). Yet, the observed lack of spatiality of quantum entities applies both to quantum particles and quantum fields, so that both ideas, of ``particle'' and ``field'', fail to properly describe a quantum entity. 

On this regard, it is sufficient to consider the situation of several quantum entities. Exactly because of the experimentally verified presence of entanglement, the so-called fields of several quantum entities are not fields that can be defined in a three-dimensional space, but in a higher dimensional configuration space. And this means that, in the same way as the quantum particles described by QM are not particles, it can similarly be claimed that quantum fields described by QFT are not fields. 

From a didactical point of view, a natural question then arises. Seeing that neither the concept of ``particle'', nor the one of ``field'' are suitable to coherently describe the nature of a non-spatial quantum entity, what would be a better, upgraded concept, in replacement of them? In other terms, what would be a more suitable conceptual pointer that we may use to designate a quantum entity?

An easy way to avoid referring to the wrong images of particles and fields is to use the Greek prefix ``pseudo'', as is already done for instance with phonons, that are not described as particles but as \emph{quasiparticles}. The prefix can be used to remind us of the fraudulence of the concept that follows it, which pretends to be something it is not. Seemingly, one could use the term of ``quantum \emph{pseudofields},'' to indicate that quantum fields, contrary to classical ones, are non-ordinary ``substances'', existing outside of our three-dimensional space. 

But there is in fact a much simpler, clearer and more elegant solution, which has been proposed by Aerts since the beginning of his research. This solution is to avoid the misleading notions of ``particle'' and ``field'' and use instead the more abstract and general notion of \emph{entity}. 

The term ``entity'' comes from the Latin \emph{entis} and refers to the essence, the beingness of something. An entity is not necessarily a spatial phenomenon, as it can also refer to mathematical, mental, conceptual aspects of our reality, and many other as well. In other terms, an entity is just a part of our multidimensional reality to which, in our role of creative observers, we are able to attach certain properties. To use Aerts' words~\cite{Aerts3}: ``An entity is a collection of properties that have a certain state of permanence to be clustered together, and a property is a state of prediction towards a certain experiment. A property, as element of the collection of properties that defines an entity, can be actual, which means that the corresponding outcome can be predicted with certainty, or potential, which means that the outcome cannot be predicted with certainty, but that the actuality of this property is available.''

What is important to observe in this general definition is that, quoting Aerts~\cite{Aerts3}: ``We drop the preconception that such clusters of properties are in space and carry a definite impact.'' Indeed, entities, be them classical, quantum or quantum-like, just need to be ``inside'' of reality, not inside our three-dimensional space. To cite once more Aerts~\cite{Aerts4}: ``Reality is not contained within space. Space is a momentaneous crystallization of a theatre for reality where the motions and interactions of the macroscopic material and energetic entities take place. But other entities - like quantum entities for example - `take place' outside space, or - and this would be another way of saying the same thing - within a space that is not the three dimensional Euclidean space.''

We would like to conclude this section expressing a symbolic ``mea culpa,'' also in the name of our colleagues, as despite the strong theoretical and experimental evidence that since a long time is under our eyes, we are still reluctant to abandon the deceptive concept o \emph{microscopic particle} - that we should at least have the decency to call \emph{quasiparticles}, as we do for instance with phonons - and replace it with the more advanced concept of \emph{quantum entity}\footnote{Of course, a similar ``mea culpa'' should be pronounced for the equally misleading concept of field. However, there are in this case, we believe, mitigating circumstances, as the concept of field, contrary to the one of particle, is usually perceived as describing a less ``tangible'' and ``visible'' aspect of reality. Also, it is often used in a figurative sense, like in the expression ``field of possibilities.'' Quantum fields are non-spatial entities, and therefore are not fields in a strict sense. Yet, they produce effects (the clicks in our counters) which are present in our three-dimensional space, and of course it is not incorrect to describe the collection of these effects (traces, clicks, etc.) as a field, namely a \emph{field of effects}. What we should bear in mind in this case, is that such a field of effects cannot capture the whole reality of the quantum entity which is at their origin, whose ``field of manifestation'' goes far beyond our restricted three-dimensional theatre.}.

The situation reminds us of when Sir Joseph John Thomson discovered that cathode rays were made of electrically charged ``corpuscles'', that later on were called electrons. His experiments showed that the so called atoms, that by definition were assumed to be uncuttable entities (according to the very etymology of their name), had in fact been ``cathodically'' cut. In other terms, all of a sudden atoms had lost their most distinctive attribute: uncuttablility! (Today we would say elementariness). But despite the experimental evidence, we did nothing to correct our nomenclature, and instead of re-baptizing the ex-atoms as molecules, and at least temporary give the name of atoms to the newly discovered electronic entities, we did nothing of the sort. 

We are not aware of the reasons for this terminological laziness, but what we can observe is that, \emph{mutatis mutandis}, the situation repeats nowadays. In fact, despite we all know that electrons neither exist in space, nor are they individual entities, we still call them particles, or corpuscles, as Thomson believed they were. And, more generally, we continue to think that the building blocks of our reality are some sort of tiny elementary particles that would describe mysterious hidden trajectories in a three-dimensional Euclidean space. Quoting Piron~\cite{Piron4}, we must admit that, for the time being ``[...] the Quantum Revolution didn't wash!''

As far as we know, our microscopic reality is not made of tiny particles moving in empty space, but of quantum entities. These entities are not three-dimensional beings, and all we concretely know about them comes from the traces they leave when they interact with our instruments. These traces are certainly present in our ordinary physical space, but in general quantum entities are not: they are not present in space, but somehow emerge (or immerge) in it every time they interact with our macroscopic three-dimensional measuring apparatus. 

Also, quantum entities generally form a single whole, and when they manifest their presence in the three-dimensional physical space, for instance in the ambit of an EPR-like experiment, they temporarily ``break up'' their wholeness, and in doing so they create correlations that violate Bell's inequalities~\cite{Aerts2}. 

Consider as a metaphorical example an elastic band entity. Imagine that the band is made of a peculiar ultra elastic material that can be stretched out on arbitrary distances, provided it is gently extended. We hang with our hand one side of the elastic and ask a colleague to hang the other side, then we walk away, putting between us and him a great spatial distance. In doing so, the elastic becomes thinner and thinner, apparently disappearing from our three-dimensional perspective. And when we are far enough, with a jerk we break the elastic, so that one fragment will shrink in our hand and the other one in the hand of our distant colleague (assuming for simplicity that the elastic can only break into two pieces). What's magic (or spooky!) about this, is that if we know the length of the unstreched elastic band (i.e., how the system was initially prepared) and measure the length of the fragment in our hand, then we can predict with certainty (by  doing a simple subtraction) the length of the fragment in the hand of our colleague, and this without having to exchange any information with him. There is no ``spooky action-at-distance,'' but only the creation of correlations in the exact moment that we break the wholeness of the elastic band into two separate pieces~\cite{Aerts2}.

\section{Conclusion}
\label{sec:8}

In the present work our principal goal was to promote a reflection in the reader as regards the subtle nature of the entities populating our microscopic (and macroscopic) reality, with a particular emphasis on the oxymoronic notion of ``microscopic particle.'' We did this by trying to bring together, in a didactical way, some of the important results, and pieces of reasoning, that were derived in the last decades by Constantin Piron, Diederik Aerts and collaborators. 

Doing this, we have also introduced our modest contribution in the longstanding effort of clarification carried out by the so-called Geneva-Brussels school in the foundations of physics: a spaghetti-entity model, proving that non-classical properties that are also non-compatible can be found as well in everyday macroscopic objects.

Of course, much more should be said on the important notion of \emph{entity}, as put forward by Aerts in his insightful creation-discovery view of reality, and we refer the reader who is not familiar with this approach to the cited articles of Aerts et al., which despite their conceptual subtleness are always written in a very clear and didactic style.

Concerning the possibility of understanding the strange quantum level of our reality, Feynman's famous saying immediately comes to mind~\cite{Feynman2}: ``I think I can safely say that nobody understands quantum mechanics.'' Considering however the number of explicit models that have been provided by Aerts and collaborators over the years, we must recognize that Feynman's admonition has today become quite obsolete.

In the present article, in addition to our newly conceived ``spaghetti model,'' we have only mentioned the ``piece of wood'' and the ``quantum machine'' examples. But there are many others. For instance, the quantum machine model has been further expanded in what is known as the ``$\epsilon$-model,''~\cite{Aerts3} allowing for the description of intermediate, more general structures, which are neither purely quantum nor purely classical. Also, Aerts did provide clarifying macroscopic examples modeling EPR non-local correlations, like his famous ``vessel of water model,''~\cite{Aerts2, Aerts5} or his ``quantum machines connected by a rigid rod.''~\cite{Aerts8} 

He provided as well very general analysis, illustrating all the subtleties inherent in our construction of reality and the relation between the modalities of such a construction and our understanding of quantum structures. We can cite, among other things, his illuminating account of the ``cracking of walnuts'' experiment~\cite{Aerts3, Aerts4} and his analysis of a human ``friendship space'', as a significant metaphor for an evolution from a non-spatial reality (our less structured micro-world) to a spatial one (our more structured macro-world)~\cite{Aerts2}.

Of course, we are not saying that, thanks to these models and the advanced conceptual views they have originated, we have now penetrated the ultimate reality of our micro-world. Such an ``ultimate reality'' level will probably remain forever hidden to our intellectual investigations. But, surely, we have pretty much advanced in revealing what our micro-world really is about, at least structurally speaking.

Having said this, we think it is relevant to conclude this article by mentioning the most recent advances of Aerts' work which, in the opinion of this author, represent the culmination of his tireless effort towards a demystified understanding of the nature of quantum entities. These advances follow from the successes of Aerts and his group, in recent years, in using the quantum formalism for the modeling of human concepts. This led him to ask a deep and thought provoking question~\cite{Aerts9}: ``If quantum mechanics as a formalism models human concepts so well, perhaps this indicates that quantum particles themselves are conceptual entities?'' 

This question was the starting point of a new interpretation of QM, which is probably today most advanced explanatory framework to understand this leading-edge theory. According to Aerts, quantum entities would~\cite{Aerts9} ``[...] interact with ordinary matter, nuclei, atoms, molecules, macroscopic material entities, measuring apparatus, ..., in a similar way to how human concepts interact with memory structures, human minds or artificial memories.'' Therefore, quoting again Aerts~\cite{Aerts9}: ``if proven correct, this new quantum interpretation would provide an explanation according to which `quantum particles' behave like something we are all very familiar with and have direct experience with, namely concepts.'' 

It is not our intention to further comment this subtle explanatory framework and its effectiveness in truly explaining phenomena such as entanglement and non-locality, which are traditionally considered (in the spirit of the above mentioned quote of Feynman)  ``not understood,'' and leave to the reader the intellectual pleasure to directly discover them in Aerts' recently published articles~\cite{Aerts9, Aerts10}. Let us just observe -- and on this we conclude -- that no doubts we have come a long way since those days when we only disposed of very na\"if images of particles and Descartian substances moving in space, to describe our truly multidimensional physical reality.

\begin{acknowledgements}

I dedicate this article to Professor Constantin Piron, one of the founders of the Geneva-Brussels school of quantum mechanics, whose I had the pleasure being the assistant about twenty years ago. Interacting with Constantin for more than a year, almost on a daily basis, irreversibly changed my way of looking to the mysteries of the quantum world, which, somehow paradoxically, used to simultaneously become less and more mysterious when I was in his presence.

The author is grateful to Diederik Aerts, for helpful communications and for his encouragement, as well as to two anonymous referees, for their careful reading of the manuscript and the many insightful comments, which have contributed in considerably improving the presentation and content of this work.

\end{acknowledgements}

\end{document}